%% file: author.tex
\documentclass[graybox]{svmult}

\usepackage{mathptmx}       
\usepackage{helvet}         
\usepackage{courier}        
\usepackage{type1cm}        
\usepackage{makeidx}         
\usepackage{graphicx}        
\usepackage{multicol}        
\usepackage[bottom]{footmisc}

\makeindex             

\begin{document}

\title*{Resolving $R_D$ and $R_{D^*}$ anomalies}
\author{Suman Kumbhakar, Ashutosh Kumar Alok, Dinesh Kumar and S Uma Sankar}
\institute{Suman Kumbhakar \at Indian Institute of Technology Bombay, Mumbai 400076, India, \email{suman@phy.iitb.ac.in}
\and Ashutosh Kumar Alok \at Indian Institute of Technology Jodhpur, Jodhpur 342011, India, \email{akalok@iitj.ac.in}
\and Dinesh Kumar \at University of Rajasthan, Jaipur 302004, India, \email{dinesh@uniraj.ac.in}
\and S Uma Sankar \at Indian Institute of Technology Bombay, Mumbai 400076, India, \email{uma@phy.iitb.ac.in}}
%
%
\maketitle


\abstract{The current world averages of the ratios $R_{D^{(*)}}$ are about $4\sigma$ away from their Standard Model prediction. These measurements indicate towards the violation of lepton flavor universality in $b\rightarrow c\,l\,\bar{\nu}$ decay.  The different new physics operators, which can explain the $R_{D^{(*)}}$ measurements, have been identified previously. We show that a simultaneous measurement of the polarization fractions of $\tau$ and $D^*$ and the angular asymmetries $A_{FB}$ and $A_{LT}$ in $B\rightarrow D^*\tau\bar{\nu}$ decay can distinguish all the new physics amplitudes and hence uniquely identify the Lorentz structure of new physics.}
\section{Introduction}
\label{sec:1}
In recent years, the evidence for charged lepton universality violation is observed in the charge current process $b\rightarrow c\tau\bar{\nu}$. The experiments, BaBar, Belle and LHCb, made several measurements of the ratios
\begin{equation}
R_{D} = \frac{\Gamma(B\rightarrow D\,\tau\,\bar{\nu})} {\Gamma(B\rightarrow D\, \{e/\mu\} \, \bar{\nu})},\hspace{0.1cm}
R_{D^*} = \frac{\Gamma(B\rightarrow D^{*}\,\tau\,\bar{\nu})} {\Gamma(B\rightarrow D^{*}\, \{e/\mu\} \, \bar{\nu})}.
\label{rdrds}
\end{equation}
The current world averges of these measurements are about $4\sigma$ away from the Standard Model (SM) predictions~\cite{Amhis:2016xyh}.

All the meson decays in eq.~(\ref{rdrds}) are driven by quark level transitions $b\rightarrow cl\bar{\nu}$. These transitions occur at tree level in the SM. The discrepancy between the measured values of $R_D$ and $R_{D^*}$ and their respective SM predictions is an indication of presence of new physics (NP) in the $b\rightarrow c\tau\bar{\nu}$ transition. The possibility of NP in $b\rightarrow c\mu\bar{\nu}$ is excluded by other data~\cite{Alok:2017qsi}. All possible NP four-Fermi operators for $b\rightarrow c\tau\bar{\nu}$ transition are listed in ref.~\cite{Freytsis:2015qca}. In ref~\cite{Alok:2017qsi}, a fit was performed between all the $b\rightarrow c\tau\bar{\nu}$ data and each of the NP interaction term. The NP terms, which can account for the all $b\rightarrow c\tau\bar{\nu}$ data, are identified and their Wilson coefficients (WCs) are calculated. It was found that there are six allowed NP solutions. Among those six solutions, four solutions are distinct with a different Lorentz structure.
In ref.~\cite{Alok:2016qyh} it was found that the tensor NP solution could be distinguished from other possibilities provided $\langle f_L\rangle$, the $D^*$ polarization fraction can be measured with an absolute uncertainty of $0.1$. 

Here, we consider four angular observables, $P_{\tau}(D^*)$ ($\tau$ polarization fraction), $f_{L}$ ($D^*$ polarization fraction), $A_{FB}$ (the forward-backward asymmetry), $A_{LT}$ (longitudinal-transverse asymmetry) in the decay $B\rightarrow D^*\tau\bar{\nu}$. Note that these asymmetries can only be measured if the momentum of the $\tau$ lepton is reconstructed. We show that a measurement of these four quantities can uniquely identify the Lorentz structure of the NP operator responsible for the present discrepancy in $R_D$ and $R_{D^*}$~\cite{Alok:2018uft}.
\section{Distinguishing different new physics solutions}
\label{sec:2}
The most general effective Hamiltonian for $b\rightarrow c\tau\bar{\nu}$ transition can be written as
\begin{equation}
H_{eff}= \frac{4 G_F}{\sqrt{2}} V_{cb}\left[O_{V_L} + \frac{\sqrt{2}}{4 G_F V_{cb}} \frac{1}{\Lambda^2} \left\lbrace \sum_i \left(C_i O_i +
 C^{'}_i O^{'}_i + C^{''}_i O^{''}_i \right) \right\rbrace \right],
\label{effH}
\end{equation}
where $G_F$ is the Fermi coupling constant, $V_{cb}$ is the Cabibbo-Kobayashi-Maskawa (CKM) 
matrix element and the NP scale $\Lambda$ is assumed to be 1 TeV. We also assume that neutrino is always left chiral. The effective Hamiltonian for the SM contains only the $O_{V_L}$ operator. The explicit forms of the four-fermion operators $O_i$, $O^{'}_i$ and $O^{''}_i$ are given in ref~\cite{Freytsis:2015qca}.
The NP effects are encoded in the NP WCs $C_i, C^{'}_i$ and $C^{''}_i$. 
Each primed and double primed operator can be expressed as a linear combination of unprimed operators through Feirz transformation.

The values of NP WCs which fit the data on the observables $R_D$, $R_{D^*}$, $R_{J/\psi}$, $P_{\tau}(D^*)$ and $\mathcal{B}(B_c\rightarrow \tau\bar{\nu})$, have been calculated previously~\cite{Alok:2017qsi}. Here $R_{J/\psi}$ is the ratio of $\mathcal{B}(B_c\rightarrow J/\psi\tau\bar{\nu})$ to $\mathcal{B}(B_c\rightarrow J/\psi\mu\bar{\nu})$~\cite{Aaij:2017tyk}. The results of these fits are listed in table~\ref{tab2}. This table also lists, for each of the NP solutions, the predicted values of the polarization fractions and the angular asymmetries in $B\rightarrow D^*\tau\bar{\nu}$ decay.
\begin{table}[htbp]
\centering 
\tabcolsep 6pt
\resizebox{\textwidth}{!}{ 
\begin{tabular}{|c|c|c|c|c|c|}
\hline\hline
   NP WCs  & Fit values  &$\langle P_{\tau}(D^*)\rangle$ & $\langle f_L\rangle$&$\langle A_{FB}\rangle$  & $\langle A_{LT}\rangle$ \\
\hline
SM  &$C_{i}=0$   &$-0.499\pm 0.004$ &$0.45\pm 0.04$& $-0.011\pm 0.007$ & $-0.245\pm 0.003$ \\
\hline 
$C_{V_L}$  & $0.149\pm 0.032$ &$-0.499\pm 0.004$&  $0.45\pm 0.04$ &$-0.011\pm 0.007$ &$-0.245\pm 0.003$  \\
\hline
$C_T$  &  $0.516 \pm 0.015$&$+0.115\pm 0.013$ & $0.14\pm 0.03$  &$-0.114\pm 0.009$& $+0.110\pm 0.009$ \\
\hline
$C''_{S_L}$ & $-0.526\pm 0.102$  &$-0.485\pm 0.003$&$0.46\pm 0.04$ &$-0.087\pm 0.011$&$-0.211\pm 0.008$ \\
\hline
$(C_{V_L},C_{V_R})$& $(-1.286, 1.512)$ & $-0.499\pm 0.004$& $0.45\pm 0.04$&$-0.371\pm 0.004$&$+0.007\pm 0.004$ \\
\hline
$(C'_{V_L},\, C'_{V_R})$  &  $(0.124, -0.058)$ & $-0.484\pm 0.005$ &$0.45\pm 0.04$& $-0.003\pm 0.007$ & $-0.243\pm 0.003$  \\
\hline
$(C''_{S_L},\, C''_{S_R})$  & $(-0.643, -0.076)$ & $-0.477\pm 0.003$&$0.46\pm 0.04$& $-0.104\pm 0.005$& $-0.202\pm 0.002$  \\
\hline\hline
\end{tabular}
}
\caption{Best fit values of NP WCs at $\Lambda=1$ TeV, taken from table IV of ref.~\cite{Alok:2017qsi}. We provide the predictions of  $\langle P_{\tau}(D^*)\rangle$, $\langle f_L\rangle$, $\langle A_{FB}\rangle$ and $\langle A_{LT}\rangle$ in decay $B\rightarrow D^*\tau\bar{\nu}$ with their uncertainties for each of the allowed solutions.  }
\label{tab2}
\end{table}
Here we compute $A_{FB}(q^2)$ and $A_{LT}(q^2)$ in $B\rightarrow D^*\tau\bar{\nu}$ decay, as functions of $q^2 = (p_B - p_{D^*})^2$, where $p_B$ and $p_{D^*}$ are the four momenta of $B$ and $D^*$ respectively.
 The predictions for $P_{\tau}(D^*)$, $f_{L}$ and $A_{FB}$ are calculated using the framework provided in \cite{Sakaki:2013bfa} and for $A_{LT}(q^2)$ we follow ref~\cite{Alok:2010zd,Duraisamy:2014sna}.

The  $B\rightarrow D^{(*)}\, l\,  \bar{\nu}$ decay distributions depend upon hadronic form-factors. The form factors for $B\rightarrow D$ decay are well known in lattice QCD \cite{Aoki:2016frl} and we use them in our analyses. For $B\rightarrow D^*$ decay, the HQET parameters are extracted using data from Belle and BaBar experiments along with lattice inputs. In this work, the numerical values of these parameters are taken from refs. \cite{Bailey:2014tva} and  \cite{Amhis:2016xyh}.

This table lists six different NP solutions but only the first four solutions are distinct~\cite{Alok:2017qsi}. Thus we have four different NP solutions with different Lorentz structures. We explore methods to distinguish between them.

\begin{figure}[t]
\centering
\resizebox{\textwidth}{!}{ 
\begin{tabular}{ccc}
\includegraphics[width=60mm]{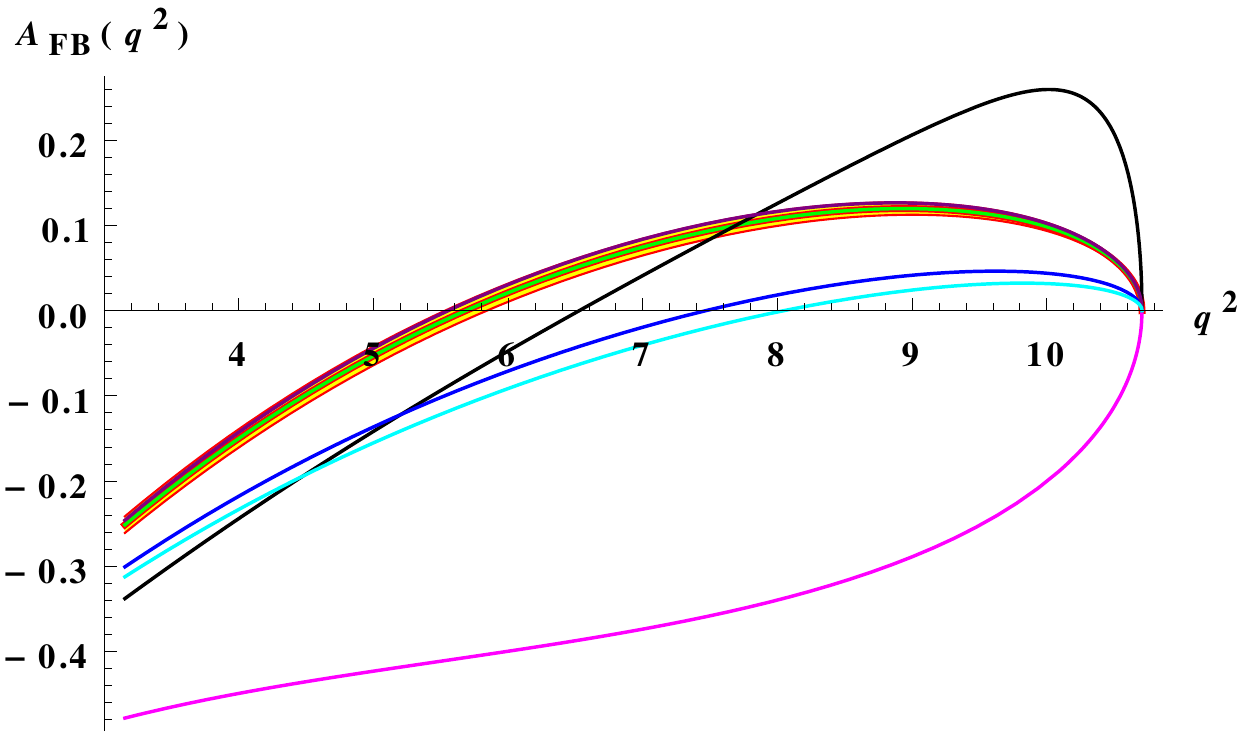}& \hspace{1cm} &
\includegraphics[width=60mm]{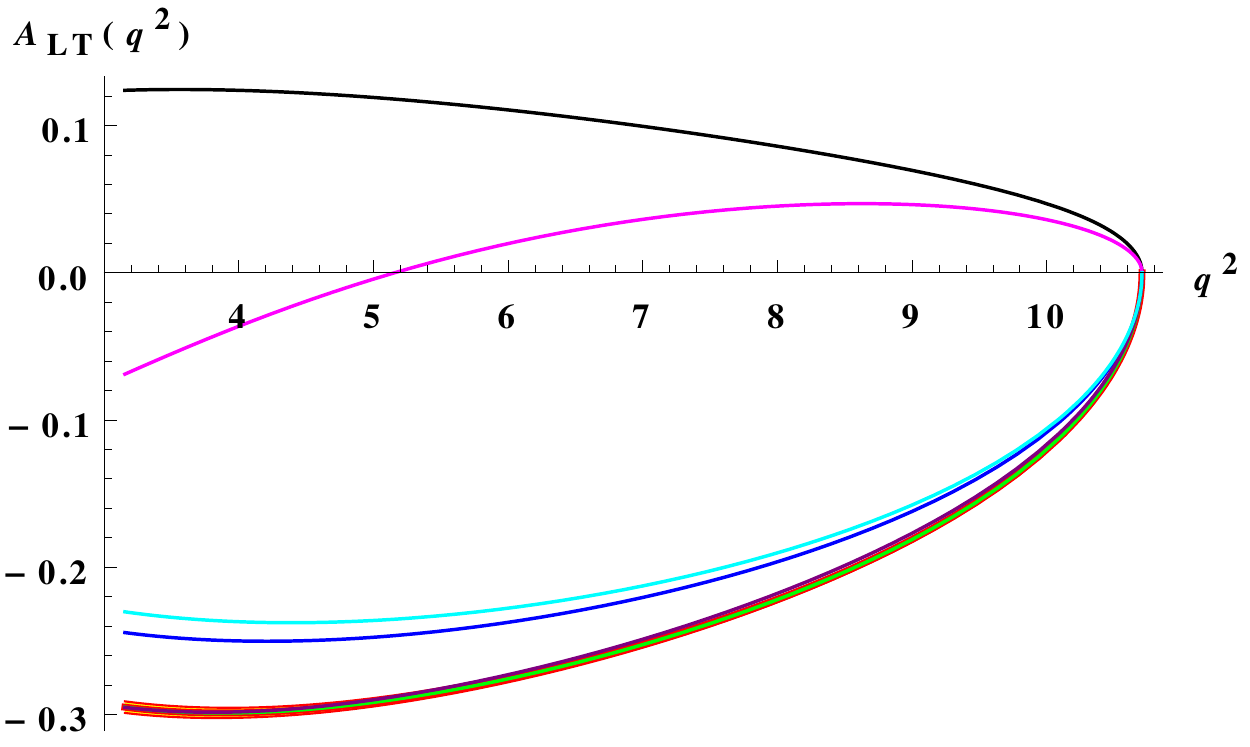} \\
\end{tabular}
}
\caption{Left and right panels correspond to $A_{FB}(q^2)$ and $A_{LT}(q^2)$, respectively for the $B\rightarrow D^*\tau\bar{\nu}$ decay. Red curves with yellow band corresponds to SM predictions. The band, representing 1$\sigma$ range, is mainly due to the uncertainties in various hadronic form factors and is obtained by adding these errors in quadrature. In each panel, the color code for the NP solutions is: $C_{V_L} = 0.149$ (green curve), $C_T = 0.516$ (black curve), $C''_{S_L} = -0.526$ (blue curve), $(C_{V_L},C_{V_R})= (-1.286,1.512)$ (magenta curve), $(C'_{V_L},C'_{V_R}) = (0.124, -0.058)$ (purple curve), $(C''_{S_L}, C''_{S_R}) = (-0.643, -0.076)$ (cyan curve).}
\label{fig1}
\end{figure}

\section{Results and Discussions}
\label{sec:3}
The average values of $P_{\tau}(D^*)$ and $f_{L}$ for all six NP solutions are given in table~\ref{tab2}. Not surprisingly, there is a large difference between the predicted values for $O_T$ solution and those for other NP solutions. If either of these observables is measured with an absolute uncertainty of $0.1$, then the $O_T$ solution is either confirmed or ruled out at $3\sigma$ level.

We now  show that the angular asymmetries $A_{FB}$ and $A_{LT}$ have a good discrimination capability between the three remaining NP WCs. The plots for $A_{FB}$ and $A_{LT}$ as a function of $q^2$ are shown in the bottom row of fig.~\ref{fig1} and their average values are listed in table~\ref{tab2}. We see that the plots of both $A_{FB}(q^2)$ and $A_{LT}(q^2)$, for $(O_{V_L}, O_{V_R})$ solution, differ significantly from the plots of all other NP solutions as do the average values. If either of these asymmetries is measured with an absolute uncertainty of $0.07$, then the $(O_{V_L}, O_{V_R})$ solution is either confirmed or ruled out at $3\sigma$ level.

So far we have identified observables which can clearly identify the $O_T$ and the $(O_{V_L}, O_{V_R})$ solutions. As we can see from table~\ref{tab2}, one needs to measure $\langle A_{FB}\rangle$ with an absolute uncertainty of $0.03$ or better to obtain a $3\sigma$ distinction between $O_{V_L}$ and $O''_{S_L}$ solutions. However, this ability to make the distinction can be improved by observing $q^2$ dependence of $A_{FB}$ for these solutions. We note that $A_{FB}(q^2)$ for $O_{V_L}$ solution has a zero crossing at $q^2 = 5.6$ GeV$^2$ whereas this crossing point occurs at $q^2 = 7.5$ GeV$^2$ for $O''_{S_L}$ solution. A calculation of $\langle A_{FB}\rangle$ in the limited range $6$ GeV$^2< q^2<q^2_{max}$ gives the result $+0.1$ for $O_{V_L}$ and $+0.01$ for $O''_{S_L}$. Hence, determining the sign of $\langle A_{FB}\rangle$, for the full $q^2$ range and for the limited higher $q^2$ range, provides a very useful tool for discrimination between these two solutions.

Hence, we find that a clear distinction can be made between the four different NP solutions to the $R_D$/$R_{D^*}$ puzzle by means of polarization fractions and angular asymmetries. 
Note that only the observables ($P_{\tau}(D^*)$ and $f_L$) isolating $O_T$ do not require the reconstruction of $\tau$ momentum. The reconstruction of $\tau$ momentum is crucial to measure the asymmetries which can distinguish between the other three NP solutions.

\input{referenc}

\end{document}

%% file: referenc.tex
%
%
%